\pdfoutput=1
\documentclass[prb,twocolumn,aps]{revtex4}
\usepackage{graphicx,graphics,color,epsfig}
\usepackage{bm}
\usepackage{amsmath}
\usepackage{amssymb}
\usepackage{epstopdf}

\makeatletter

\newcommand{\Rmnum}[1]{\expandafter\@slowromancap\romannumeral #1@}
\newcommand*{\rom}[1]{\expandafter\@slowromancap\romannumeral #1@}
\makeatother

\usepackage{color}

\renewcommand{\Re}{\operatorname{Re}}
\renewcommand{\Im}{\operatorname{Im}}

\begin{document}
\title{Theoretical study of the electronic correlation and superconducting pairing in La$_{2.85}$Pr$_{0.15}$Ni$_2$O$_7$ film grown on SrLaAlO$_4$}

\author{Yi Gao}
\email{flygaoonly@njnu.edu.cn}
\affiliation{Center for Quantum Transport and Thermal Energy Science, Jiangsu Key Lab on Opto-Electronic Technology, School of Physics and Technology, Nanjing Normal University, Nanjing 210023, China}

\begin{abstract}
We investigate the electronic correlation effects in La$_{2.85}$Pr$_{0.15}$Ni$_2$O$_7$ film grown on SrLaAlO$_4$ by fluctuation-exchange approximation. The $\gamma$ and $\delta$ bands are found to shift down with correlation while their Fermi surfaces become highly damped and cannot to be resolved experimentally. In contrast, the $\alpha$ and $\beta$ Fermi surfaces do not vary much with correlation and they are sharp enough to be detected in experiments. The spin susceptibility peaks at a wave vector in the odd channel, connecting the symmetric $\gamma$ band and the asymmetric $\delta$ band. The superconducting pairing symmetry is robustly $s$-wave and the $\delta$ band has the largest pairing magnitude. All the findings suggest a dominant role of the $3d_{z^{2}}$ orbital in this material.
\end{abstract}

\maketitle

\section{introduction}
In 2023, superconductivity was discovered in La$_3$Ni$_2$O$_7$ under high pressure, with a transition temperature $T_c\approx80$ K \cite{discovery}. In the high-pressure phase, it features an orthorhombic structure of $Fmmm$ space group. The Ni atoms form a bilayer square lattice with the 3$d_{x^2-y^2}$ and 3$d_{z^2}$ orbitals of Ni dominating the energy bands close to the Fermi level. However, the high-pressure condition that is necessary for superconductivity is incompatible with photoemission experimental techniques, leaving the electronic structure of the superconducting state undetermined experimentally.

Recently, ambient-pressure superconductivity with $T_c\approx40$ K was reported in bilayer nickelate ultrathin films \cite{thinfilm1,thinfilm2}, offering an opportunity for investigating their electronic structures by angle-resolved photoemission spectroscopy (ARPES). Several ARPES measurements have been performed in these thin films \cite{arpes_la2prni2o7,arpes_la2.85pr0.15ni2o7,arpes_(laprsm)3ni2o7,arpes_la2.79sr0.21ni2o7}. The common feature is, the 3$d_{x^2-y^2}$ orbital derived $\alpha$ and $\beta$ Fermi surfaces can be clearly identified. In contrast, the location of the 3$d_{z^2}$ orbital derived $\gamma$ band is under debate, there are conflicting conclusions on whether it crosses the Fermi level or not \cite{arpes_la2prni2o7,arpes_la2.85pr0.15ni2o7,arpes_la2.79sr0.21ni2o7}. Even if it forms a Fermi surface, the Fermi surface is highly damped with a low spectral weight \cite{arpes_la2.85pr0.15ni2o7,arpes_(laprsm)3ni2o7}. In addition, little attention has been paid to another 3$d_{z^2}$ orbital derived $\delta$ band.

The superconducting pairing symmetry has been investigated extensively in both bulk La$_3$Ni$_2$O$_7$ under high pressure and thin films at ambient pressure. It is found that the $s_{\pm}$-wave pairing symmetry, which mainly stems from the inter-layer pairing in the 3$d_{z^2}$ orbital \cite{s1,s7,s8,s9,s11,s12}, strongly competes with the $d_{x^2-y^2}$-wave \cite{sd1,sd2,sd3,sd6,d1,d3} and the $d_{xy}$-wave ones \cite{sd4,sd5,d2,d3}. The latter two $d$-wave symmetries are originated predominantly from the 3$d_{x^2-y^2}$ orbital \cite{sd2,sd3,sd6,sd1,d1}.

In this work, we study the electronic correlation effects in La$_{2.85}$Pr$_{0.15}$Ni$_2$O$_7$ film grown on SrLaAlO$_4$ by fluctuation-exchange (FLEX) approximation \cite{junhuazhang}, since the ARPES measurement has suggested a considerable band renormalization from the density functional theory \cite{arpes_la2.85pr0.15ni2o7}. We found that, the 3$d_{z^2}$-orbital bands are more affected by correlation and their Fermi surfaces become highly damped and cannot be resolved, in qualitative agreement with experiment. The spin susceptiblity peaks at a wave vector connecting two 3$d_{z^2}$-orbital bands and the superconducting pairing is predominatly in the 3$d_{z^2}$ orbital. All the findings indicate the 3$d_{z^2}$ orbital plays the most important role in the bilayer nickelate superconductors.

\section{method}

We adopt a bilayer two-orbital model of La$_{2.85}$Pr$_{0.15}$Ni$_2$O$_7$ film grown on SrLaAlO$_4$ \cite{bilayermodel}. The two orbitals 3$d_{x^2-y^2}$ and 3$d_{z^2}$ of Ni are denoted as $x$ and $z$, respectively. The tight-binding part of the Hamiltonian can be written as $H_0=\sum_{\mathbf{k}\sigma}\psi_{\mathbf{k}\sigma}^{\dag}M_{\mathbf{k}}\psi_{\mathbf{k}\sigma}$, where $\psi_{\mathbf{k}\sigma}^{\dag}=(c_{\mathbf{k}1x\sigma}^{\dag},c_{\mathbf{k}2x\sigma}^{\dag},c_{\mathbf{k}1z\sigma}^{\dag},c_{\mathbf{k}2z\sigma}^{\dag})$ and
\begin{eqnarray}
\label{h0}
M_{\mathbf{k}}&=&\begin{pmatrix}
T_{\mathbf{k}}^{x}&T_{\mathbf{k}}^{'x}&V_{\mathbf{k}}&V_{\mathbf{k}}^{'}\\
T_{\mathbf{k}}^{'x}&T_{\mathbf{k}}^{x}&V_{\mathbf{k}}^{'}&V_{\mathbf{k}}\\
V_{\mathbf{k}}&V_{\mathbf{k}}^{'}&T_{\mathbf{k}}^{z}&T_{\mathbf{k}}^{'z}\\
V_{\mathbf{k}}^{'}&V_{\mathbf{k}}&T_{\mathbf{k}}^{'z}&T_{\mathbf{k}}^{z}
\end{pmatrix}.
\end{eqnarray}
Here $c_{\mathbf{k}lo\sigma}^{\dag}$ creates a spin $\sigma$ ($\sigma=\uparrow,\downarrow$) electron with momentum $\mathbf{k}$ in the layer $l$ ($l=1,2$) and orbital $o$ ($o=x,z$). In addition,
\begin{eqnarray}
T_{\mathbf{k}}^{x/z}&=&2t_{1}^{x/z}(\cos k_x+\cos k_y)+4t_{2}^{x/z}\cos k_x\cos k_y\nonumber\\
&+&2t_{4}^{x/z}(\cos 2k_x+\cos 2k_y)\nonumber\\
&+&2t_{5}^{x/z}(\cos 3k_x+\cos 3k_y)+\epsilon^{x/z},\nonumber\\
T_{\mathbf{k}}^{'x/z}&=&t_{\bot}^{x/z}+2t_{3}^{x/z}(\cos k_x+\cos k_y),\nonumber\\
V_{\mathbf{k}}&=&2t_{3}^{xz}(\cos k_x-\cos k_y)+2t_{5}^{xz}(\cos 2k_x-\cos 2k_y),\nonumber\\
V_{\mathbf{k}}^{'}&=&2t_{4}^{xz}(\cos k_x-\cos k_y).
\end{eqnarray}

By defining
\begin{eqnarray}
\label{symmetric_antisymmetric}
d_{\mathbf{k}So\sigma}&=&\frac{1}{\sqrt{2}}(c_{\mathbf{k}1o\sigma}+c_{\mathbf{k}2o\sigma}),\nonumber\\
d_{\mathbf{k}Ao\sigma}&=&\frac{1}{\sqrt{2}}(c_{\mathbf{k}1o\sigma}-c_{\mathbf{k}2o\sigma}),
\end{eqnarray}
the tight-binding Hamiltonian can be written in the symmetric and asymmetric basis with respect to exchanging the two layer indices as $H_0=\sum_{\mathbf{k}\sigma}\varphi_{\mathbf{k}\sigma}^{\dag}M^{'}_{\mathbf{k}}\varphi_{\mathbf{k}\sigma}$, where $\varphi_{\mathbf{k}\sigma}^{\dag}=(d_{\mathbf{k}Sx\sigma}^{\dag},d_{\mathbf{k}Sz\sigma}^{\dag},d_{\mathbf{k}Ax\sigma}^{\dag},d_{\mathbf{k}Az\sigma}^{\dag})$ and
\begin{eqnarray}
\label{M'k}
M^{'}_{\mathbf{k}}&=&\begin{pmatrix}
T_{\mathbf{k}}^{x}+T_{\mathbf{k}}^{'x}&V_{\mathbf{k}}+V_{\mathbf{k}}^{'}&0&0\\
V_{\mathbf{k}}+V_{\mathbf{k}}^{'}&T_{\mathbf{k}}^{z}+T_{\mathbf{k}}^{'z}&0&0\\
0&0&T_{\mathbf{k}}^{x}-T_{\mathbf{k}}^{'x}&V_{\mathbf{k}}-V_{\mathbf{k}}^{'}\\
0&0&V_{\mathbf{k}}-V_{\mathbf{k}}^{'}&T_{\mathbf{k}}^{z}-T_{\mathbf{k}}^{'z}
\end{pmatrix}.\nonumber\\
\end{eqnarray}
$M^{'}_{\mathbf{k}}$ is block-diagonalized and its eigenvalues can be expressed as

\begin{eqnarray}
\label{eigenvalues}
E_{\mathbf{k}}^{\gamma}&=&\frac{1}{2}\{T_{\mathbf{k}}^{x}+T_{\mathbf{k}}^{'x}+T_{\mathbf{k}}^{z}+T_{\mathbf{k}}^{'z}\nonumber\\
&-&[(T_{\mathbf{k}}^{x}+T_{\mathbf{k}}^{'x}-T_{\mathbf{k}}^{z}-T_{\mathbf{k}}^{'z})^2+4(V_{\mathbf{k}}+V_{\mathbf{k}}^{'})^2]^{\frac{1}{2}}\},\nonumber\\
E_{\mathbf{k}}^{\alpha}&=&\frac{1}{2}\{T_{\mathbf{k}}^{x}+T_{\mathbf{k}}^{'x}+T_{\mathbf{k}}^{z}+T_{\mathbf{k}}^{'z}\nonumber\\
&+&[(T_{\mathbf{k}}^{x}+T_{\mathbf{k}}^{'x}-T_{\mathbf{k}}^{z}-T_{\mathbf{k}}^{'z})^2+4(V_{\mathbf{k}}+V_{\mathbf{k}}^{'})^2]^{\frac{1}{2}}\},\nonumber\\
E_{\mathbf{k}}^{\beta}&=&\frac{1}{2}\{T_{\mathbf{k}}^{x}-T_{\mathbf{k}}^{'x}+T_{\mathbf{k}}^{z}-T_{\mathbf{k}}^{'z}\nonumber\\
&-&[(T_{\mathbf{k}}^{x}-T_{\mathbf{k}}^{'x}-T_{\mathbf{k}}^{z}+T_{\mathbf{k}}^{'z})^2+4(V_{\mathbf{k}}-V_{\mathbf{k}}^{'})^2]^{\frac{1}{2}}\},\nonumber\\
E_{\mathbf{k}}^{\delta}&=&\frac{1}{2}\{T_{\mathbf{k}}^{x}-T_{\mathbf{k}}^{'x}+T_{\mathbf{k}}^{z}-T_{\mathbf{k}}^{'z}\nonumber\\
&+&[(T_{\mathbf{k}}^{x}-T_{\mathbf{k}}^{'x}-T_{\mathbf{k}}^{z}+T_{\mathbf{k}}^{'z})^2+4(V_{\mathbf{k}}-V_{\mathbf{k}}^{'})^2]^{\frac{1}{2}}\}.\nonumber\\
\end{eqnarray}
Therefore we get two symmetric bands $\alpha$ and $\gamma$, as well as two asymmetric ones $\beta$ and $\delta$. The bare Green's function matrix is defined as $G_0(k)=(ip_nI-M_\mathbf{k})^{-1}$. Here $I$ is the unit matrix and $k=(\mathbf{k},ip_n)$, with $p_n=(2n-1)\pi T$ and $T$ being the temperature.


The multiorbital Hubbard interaction is written as \cite{scalapino,kubo}
\begin{eqnarray}
\label{honsite}
H_{int}&=&\sum_{i}\sum_{l}\Bigg[U\sum_{o}n_{ilo\uparrow}n_{ilo\downarrow}
+(U^{'}-\frac{J_H}{2})\sum_{o>o^{'}}n_{ilo}n_{ilo^{'}}\nonumber\\
&-&J_H\sum_{o>o^{'}}2\mathbf{S}_{ilo}\cdot\mathbf{S}_{ilo^{'}}
+J^{'}\sum_{o\neq o^{'}}c_{ilo\uparrow}^{\dag}c_{ilo\downarrow}^{\dag}c_{ilo^{'}\downarrow}c_{ilo^{'}\uparrow}\Bigg]\nonumber\\
&=&\sum_{i}\sum_{l}\Bigg[U\sum_{o}n_{ilo\uparrow}n_{ilo\downarrow}
+U^{'}\sum_{o>o^{'}}n_{ilo}n_{ilo^{'}}\nonumber\\
&+&J_H\sum_{o>o^{'}}\sum_{\sigma \sigma^{'}}c_{ilo\sigma}^{\dag}c_{ilo^{'}\sigma^{'}}^{\dag}c_{ilo\sigma^{'}}c_{ilo^{'}\sigma}\nonumber\\
&+&J^{'}\sum_{o\neq o^{'}}c_{ilo\uparrow}^{\dag}c_{ilo\downarrow}^{\dag}c_{ilo^{'}\downarrow}c_{ilo^{'}\uparrow}\Bigg],
\end{eqnarray}
where $U$, $U^{'}$, $J_H$ and $J^{'}$ are the strength of intra-orbital Coulomb interaction, inter-orbital Coulomb interaction, Hund's coupling and pair hopping, respectively.

We follow the standard FLEX procedure \cite{junhuazhang}. The bare susceptibility $\chi_{0}(q)$ is
\begin{eqnarray}
\label{X0}
&&\chi_{0}^{s_1s_4,s_2s_3}(q)
=\chi_{0}^{l_1o_1l_4o_4,l_2o_2l_3o_3}(q)\nonumber\\
&=&\frac{1}{2N}\int_{0}^{1/T}d\tau e^{i\omega_n\tau}\sum_{\mathbf{k},\mathbf{k}^{'},\sigma,\sigma^{'}}\nonumber\\
&&\langle T_{\tau}c_{\mathbf{k}+\mathbf{q}l_1o_1\sigma}^{\dag}(\tau)c_{\mathbf{k}l_4o_4\sigma}(\tau) c_{\mathbf{k}^{'}-\mathbf{q}l_2o_2\sigma^{'}}^{\dag}(0)c_{\mathbf{k}^{'}l_3o_3\sigma^{'}}(0)\rangle\nonumber\\
&=&-\frac{T}{N}\sum_k G^{s_2s_4}(k+q)G^{s_1s_3}(k).
\end{eqnarray}
Here $s_i=(l_i,o_i)$ is the combined layer and orbital index with $i=1,\ldots,4$. $N$ is the number of unit cells and $q=(\mathbf{q},i\omega_n)$, with $\omega_n=2n\pi T$. $G(k)$ is the renormalized Green's function matrix to be determined self-consistently.

Within the FLEX approximation, the spin and charge susceptibilities can be written as
\begin{eqnarray}
\label{RPA}
\chi_{s}(q)&=&[I-\chi_0(q)U_s]^{-1}\chi_0(q),\nonumber\\
\chi_{c}(q)&=&[I+\chi_0(q)U_c]^{-1}\chi_0(q),
\end{eqnarray}
where the nonzero matrix elements of $U_s$ and $U_c$ are
\begin{eqnarray}
\label{Us}
U_{s}^{lo_1 lo_2,lo_3 lo_4}&=&\begin{cases}
U&\text{$o_1=o_2=o_3=o_4$},\\
U^{'}&\text{$o_1=o_4\neq o_3=o_2$},\\
J_H&\text{$o_1=o_2\neq o_3=o_4$},\\
J^{'}&\text{$o_1=o_3\neq o_2=o_4$},
\end{cases}
\end{eqnarray}
and
\begin{eqnarray}
\label{Uc}
U_{c}^{lo_1 lo_2,lo_3 lo_4}&=&\begin{cases}
U&\text{$o_1=o_2=o_3=o_4$},\\
-U^{'}+2J_H&\text{$o_1=o_4\neq o_3=o_2$},\\
2U^{'}-J_H&\text{$o_1=o_2\neq o_3=o_4$},\\
J^{'}&\text{$o_1=o_3\neq o_2=o_4$},
\end{cases}\nonumber\\
\end{eqnarray}
with $l$ being the layer index and $o_1,o_2,o_3,o_4$ being the orbital indices.

The normal self energy is
\begin{eqnarray}
\label{normalselfenergy}
\Sigma^{s_1s_2}(k)&=&\frac{T}{N}\sum_{q}\sum_{s_3s_4}G^{s_3s_4}(k-q)V_n^{s_2s_4,s_3s_1}(q),\nonumber\\
\end{eqnarray}
where
\begin{eqnarray}
\label{Vn}
V_n(q)&=&\frac{3}{2}U_s\chi_s(q)U_s+\frac{1}{2}U_c\chi_c(q)U_c\nonumber\\
&-&\frac{3}{4}U_s\chi_0(q)U_s-\frac{1}{4}U_c\chi_0(q)U_c\nonumber\\
&+&\frac{1}{2}(3U_s-U_c).
\end{eqnarray}
Then the renormalized Green's function is
\begin{eqnarray}
\label{renormalizedGreenfunction}
G(k)&=&\{G_0^{-1}(k)-\Sigma(k)+\mu I\}^{-1},
\end{eqnarray}
where $\mu$ is the chemical potential. The electron occupation is
\begin{eqnarray}
n&=&\frac{1}{2N}\sum_{\mathbf{k},l,o,\sigma}\langle c_{\mathbf{k}lo\sigma}^{\dag}c_{\mathbf{k}lo\sigma}\rangle\nonumber\\
&=&\frac{1}{N}\sum_{\mathbf{k},s}G^{ss}(\mathbf{k},-0^{+})\nonumber\\
&=&\frac{T}{N}\sum_{\mathbf{k},s,p_n}G^{ss}(\mathbf{k},ip_n)e^{ip_n0^{+}}.\nonumber\\
\end{eqnarray}
Then by using the relation \cite{mahan}
\begin{eqnarray}
T\sum_{p_n}G^{ss}(\mathbf{k},ip_n)e^{ip_n0^{+}}&=&\frac{1}{2}+T\sum_{p_n}G^{ss}(\mathbf{k},ip_n),\nonumber\\
\end{eqnarray}
we have
\begin{eqnarray}
\label{n}
n&=&\frac{1}{2N}\sum_{\mathbf{k},s}1+
\frac{T}{N}\sum_{\mathbf{k},s,p_n}G^{ss}(\mathbf{k},ip_n)\nonumber\\
&=&2+\frac{T}{N}\sum_{k,s}G^{ss}(k).
\end{eqnarray}

Equations (\ref{X0}) to (\ref{n}) are solved self-consistently by iteration, while convergence is reached when the absolute error of $\Sigma(k)$ is less than $10^{-3}$. At the end of each iteration, the chemical potential $\mu$ is adjusted to keep the electron occupation $n$ fixed, according to Eqs. (\ref{renormalizedGreenfunction}) and (\ref{n}).

Once the renormalized Green's function is obtained,
close to $T_c$, the linearized Eliashberg equation can be expressed as \cite{eliashberg}
\begin{eqnarray}
\label{Eliashberg}
\lambda \phi^{s_3s_4}(k)&=&-\frac{T}{N}\sum_q\sum_{s_1,s_2,s_5,s_6} G^{s_1s_2}(k-q)G^{s_6s_5}(q-k)\nonumber\\
&&V_{a}^{s_6s_4,s_1s_3}(q)\phi^{s_2s_5}(k-q),
\end{eqnarray}
where $\phi(k)$ is the anomalous self energy and the spin-singlet pairing interaction is \cite{eliashberg,junhuazhang}
\begin{eqnarray}
\label{V}
V_a(q)&=&
\frac{1}{2}[3U_s\chi_{s}(q)U_s-U_c\chi_{c}(q)U_c+U_s+U_c].\nonumber\\
\end{eqnarray}

Equation (\ref{Eliashberg}) is solved by the power method \cite{power method} to find the largest positive eigenvalue $\lambda$ and the corresponding $\phi(k)$ is the preferred pairing function.
In the iterative process, due to the anti-commutation relation of the fermions, the initial input $\phi(k)$ should satisfy \cite{oddfrequency}
\begin{eqnarray}
\label{anticommutation}
\phi^{s_1s_2}(k)&=&
\phi^{s_2s_1}(-k),
\end{eqnarray}
for spin singlet pairing.

Throughout this work, the number of unit cells is $N=64\times64$ and the tight-binding parameters are \cite{bilayermodel}
\begin{eqnarray}
t_{1}^{x/z}&=&-0.466/-0.126,t_{2}^{x/z}=0.062/-0.016,\nonumber\\
t_{3}^{x/z}&=&-0.001/0.033,t_{4}^{x/z}=-0.064/-0.014,\nonumber\\
t_{5}^{x/z}&=&-0.015/-0.003,\epsilon^{x/z}=0.87/0.351,\nonumber\\
t_{\bot}^{x/z}&=&0.001/-0.439,t_{3}^{xz}=0.229,\nonumber\\
t_{4}^{xz}&=&-0.032,t_{5}^{xz}=0.026,
\end{eqnarray}
in units of eV. The summations over momentum and frequency in Eqs. (\ref{X0}), (\ref{normalselfenergy}) and (\ref{Eliashberg}) are done by fast Fourier transformation, where we use $8192$ Matsubara frequencies ($-8191\pi T\leq p_n\leq8191\pi T$ and $-8190\pi T\leq\omega_n\leq8192\pi T$). The temperature is set to be $T=0.004$ eV ($T\approx46$ K) and the electron occupation is fixed at $n=1.3$, since the ARPES measurement has suggested a $0.2$ per Ni hole doping relative to the parent bulk \cite{arpes_la2.85pr0.15ni2o7}. The interaction strength in Eq. (\ref{honsite}) satisfies $U^{'}=U-2J_H$ and we fix $J_H=J^{'}=0.56$ eV \cite{bilayermodel}. The Green's function $G(\mathbf{k},ip_n)$ and the normal self energy $\Sigma(\mathbf{k},ip_n)$ are analytically continued to $G(\mathbf{k},\omega+i\eta)$ and $\Sigma(\mathbf{k},\omega+i\eta)$, respectively, by a $1024$-point Pad$\acute{e}$ approximation \cite{pade} and $\eta$ is set to be $\pi T$. The spectral function measured by ARPES is calculated as
\begin{eqnarray}
A(\mathbf{k},\omega)&=&-\frac{2}{\pi}\sum_{s}\Im G^{ss}(\mathbf{k},\omega+i\eta).
\end{eqnarray}
Here $\Im$ denotes the imaginary part.

\section{results and discussion}

\begin{figure}
\includegraphics[width=1\linewidth]{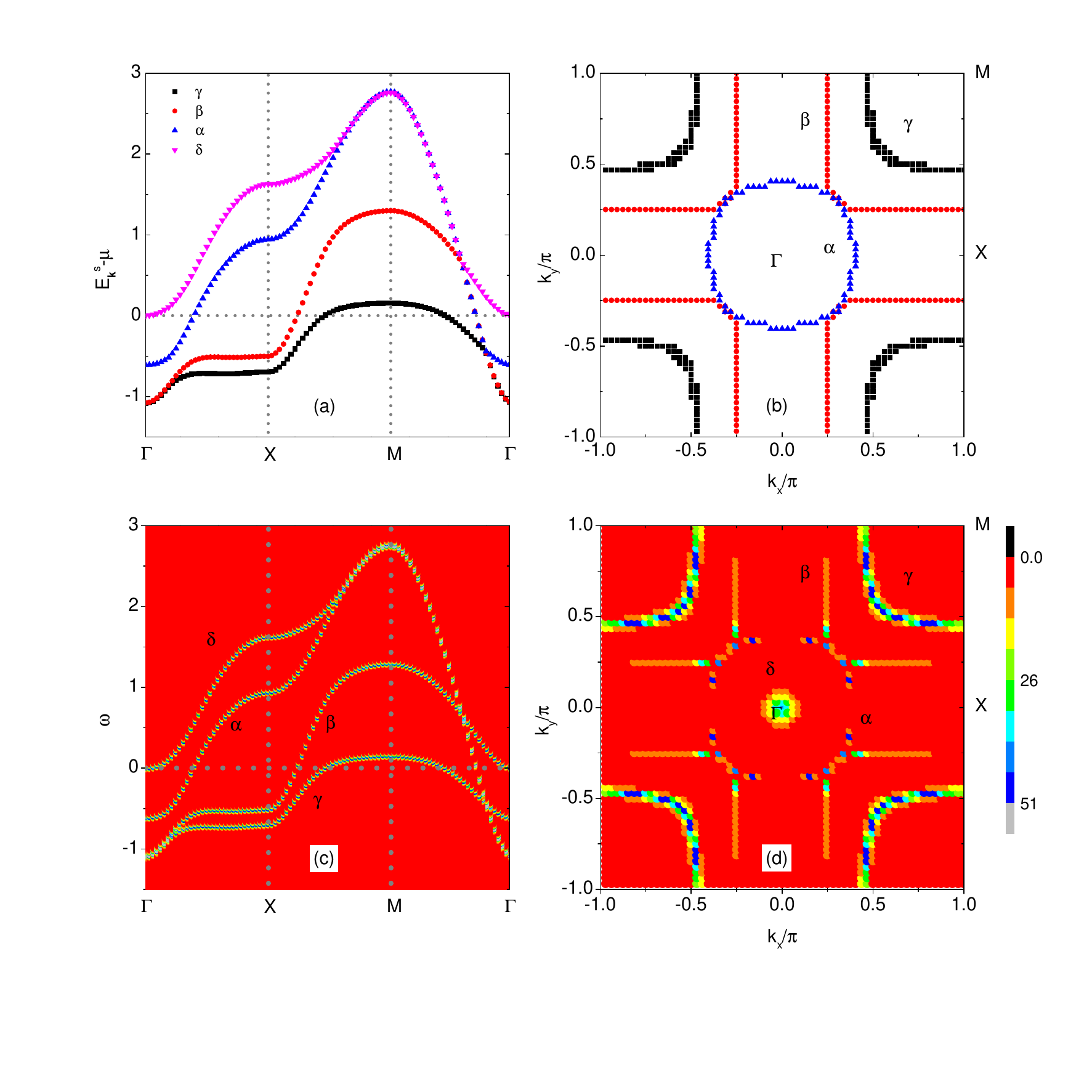}
 \caption{\label{band_structure} (a) The band dispersion of Eq. (\ref{eigenvalues}) along the high-symmetry directions, measured from the chemical potential. (b) The corresponding Fermi surface. (c) $A(\mathbf{k},\omega)$ along the high-symmetry directions. (d) $A(\mathbf{k},0)$. Here $U=J_H=0$.}
\end{figure}

First we show the band structure in the absence of interaction. Figure \ref{band_structure}(a) plots the band dispersion of Eq. (\ref{eigenvalues}), measured from the chemical potential. $\alpha$, $\beta$ and $\gamma$ bands intersect the Fermi level and form three Fermi surfaces shown in Fig. \ref{band_structure}(b). The bottom of the $\delta$ band is located at $\Gamma$ and $E_{\Gamma}^{\delta}-\mu\approx0.002$. Therefore the $\delta$ band is slightly above the Fermi level and does not form Fermi surface. The spectral functions plotted in Figs. \ref{band_structure}(c) and \ref{band_structure}(d) agree with those in Figs. \ref{band_structure}(a) and \ref{band_structure}(b), respectively. In Fig. \ref{band_structure}(d), the high-intensity spot at $\Gamma$ is due to the $\delta$ band slightly above the Fermi level, leading to a leakage of intensity to $\omega=0$.

\begin{figure*}
\includegraphics[width=1\linewidth]{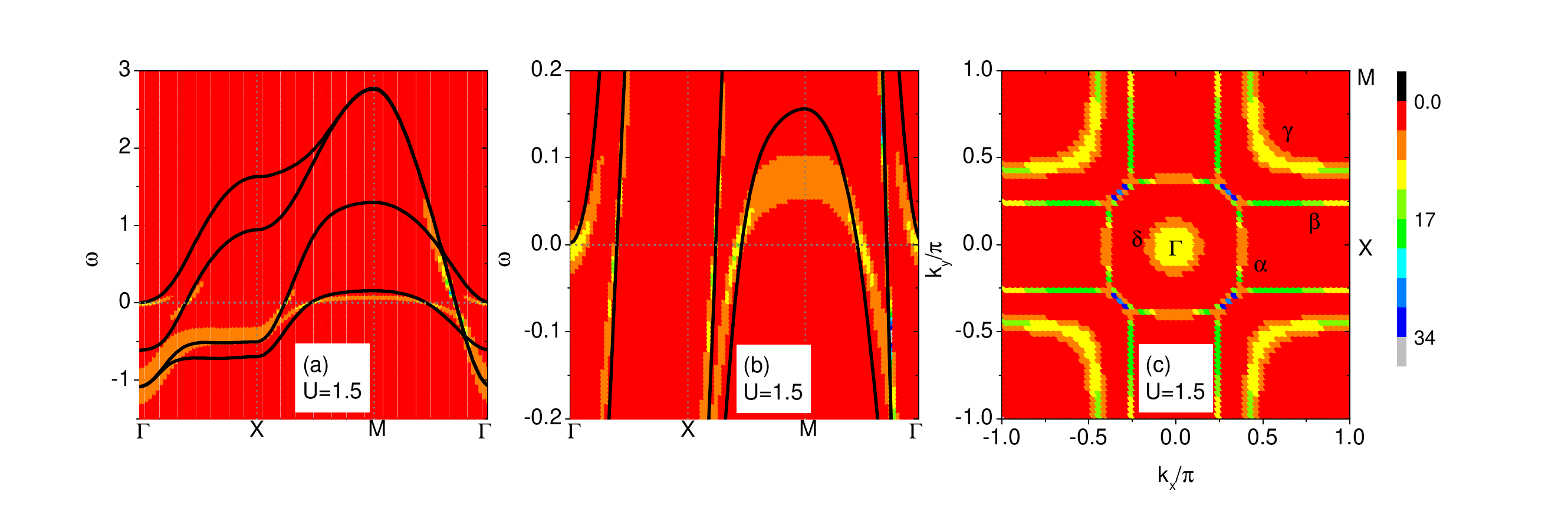}
 \caption{\label{band_structure_u=1.5} (a) $A(\mathbf{k},\omega)$ along the high-symmetry directions. (b) Similar to (a), but in a zoomed-in energy window of $-0.2<\omega<0.2$. (c) $A(\mathbf{k},0)$. Here $U=1.5$. The black lines in (a) and (b) denote the bare bands.}
\end{figure*}

\begin{figure*}
\includegraphics[width=1\linewidth]{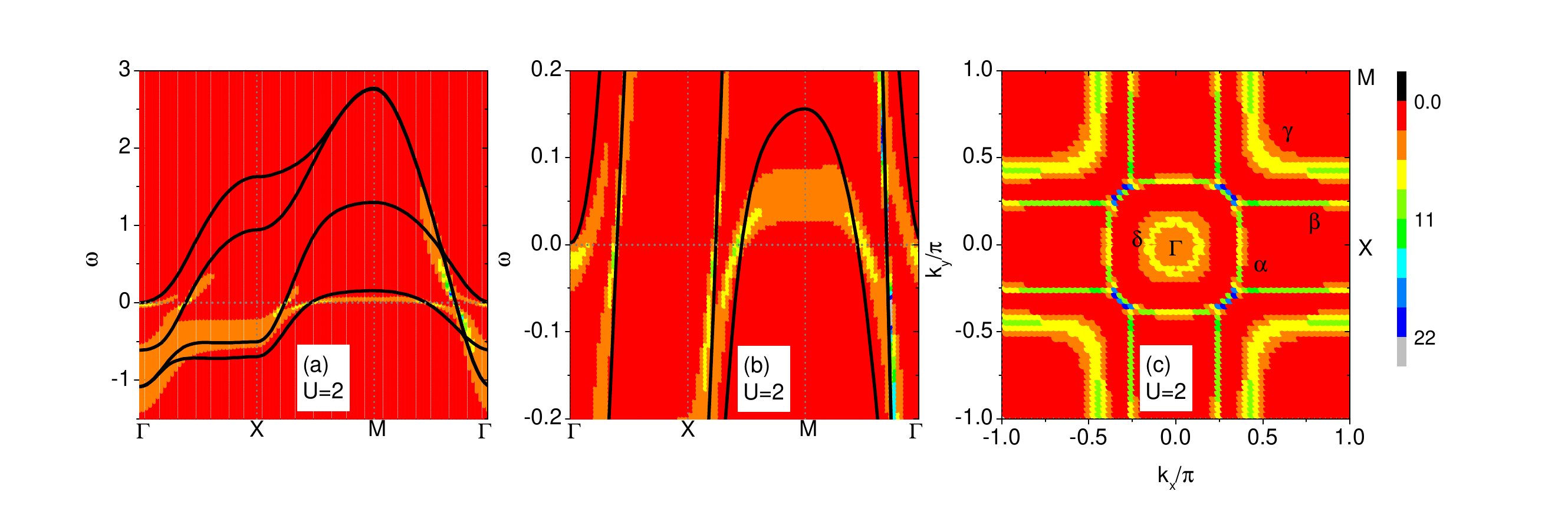}
 \caption{\label{band_structure_u=2} Similar to Fig. \ref{band_structure_u=1.5}, but at $U=2$.}
\end{figure*}

\begin{figure*}
\includegraphics[width=1\linewidth]{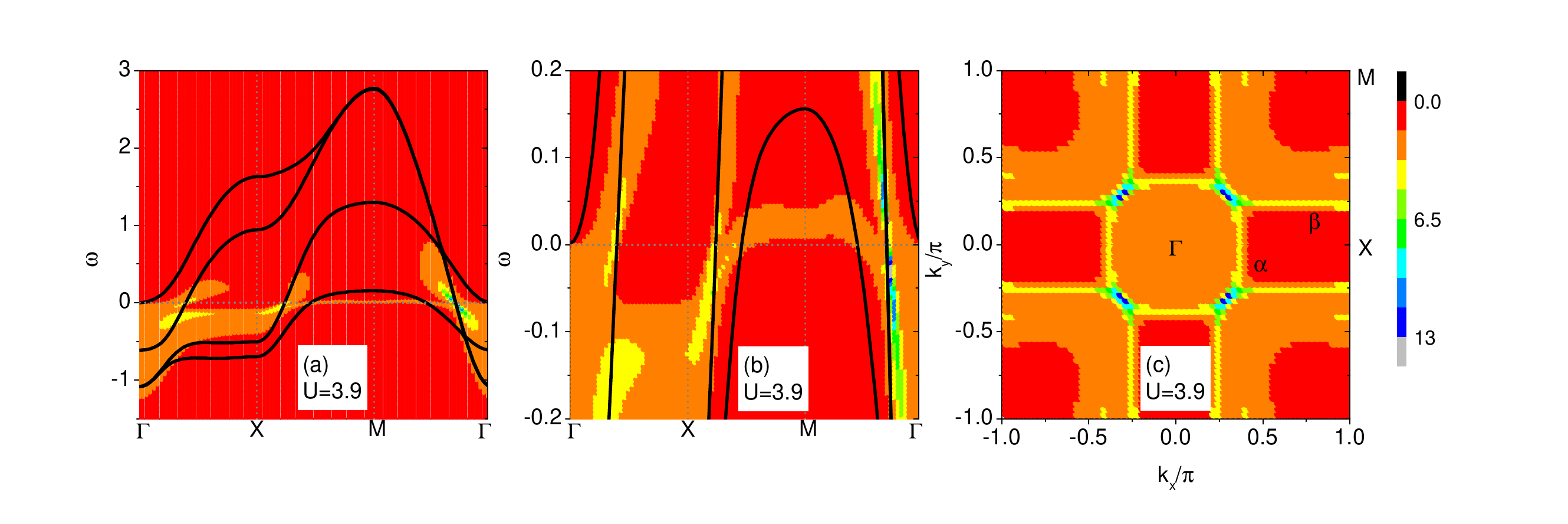}
 \caption{\label{band_structure_u=3.9} Similar to Fig. \ref{band_structure_u=1.5}, but at $U=3.9$.}
\end{figure*}

Then we study the band renormalization in the presence of interaction. At $U=1.5$, the spectral function in Fig. \ref{band_structure_u=1.5}(a) suggests that the bands become highly damped, while they are coherent only in the vicinity of zero energy, as shown in Fig. \ref{band_structure_u=1.5}(b). Both the bottom of the $\delta$ band at $\Gamma$ and the top of the $\gamma$ band at $M$ shift downward and the Fermi velocities of all the four bands decrease. The Fermi surface in Fig. \ref{band_structure_u=1.5}(c) shows that the $\alpha$, $\beta$ and $\gamma$ Fermi surfaces vary little compared to the bare ones in Fig. \ref{band_structure}(d), while the high-intensity spot at $\Gamma$ becomes larger since the $\delta$ band moves down and crosses the Fermi level there. As $U$ increases to $2$, the $\delta$ band at $\Gamma$ moves down further, as shown in Figs. \ref{band_structure_u=2}(a) and \ref{band_structure_u=2}(b). Correspondingly, a clear $\delta$ Fermi surface can be resolved around $\Gamma$ in Fig. \ref{band_structure_u=2}(c). Finally, when $U$ increases to $3.9$, the $\delta$ band around $\Gamma$ becomes highly damped and the $\gamma$ band around $M$ becomes almost flat and is also highly damped, as shown in Figs. \ref{band_structure_u=3.9}(a) and \ref{band_structure_u=3.9}(b). Therefore, there are no clear $\delta$ and $\gamma$ Fermi surfaces and only the $\alpha$ and $\beta$ ones can be resolved in Fig. \ref{band_structure_u=3.9}(c).

\begin{figure}
\includegraphics[width=1\linewidth]{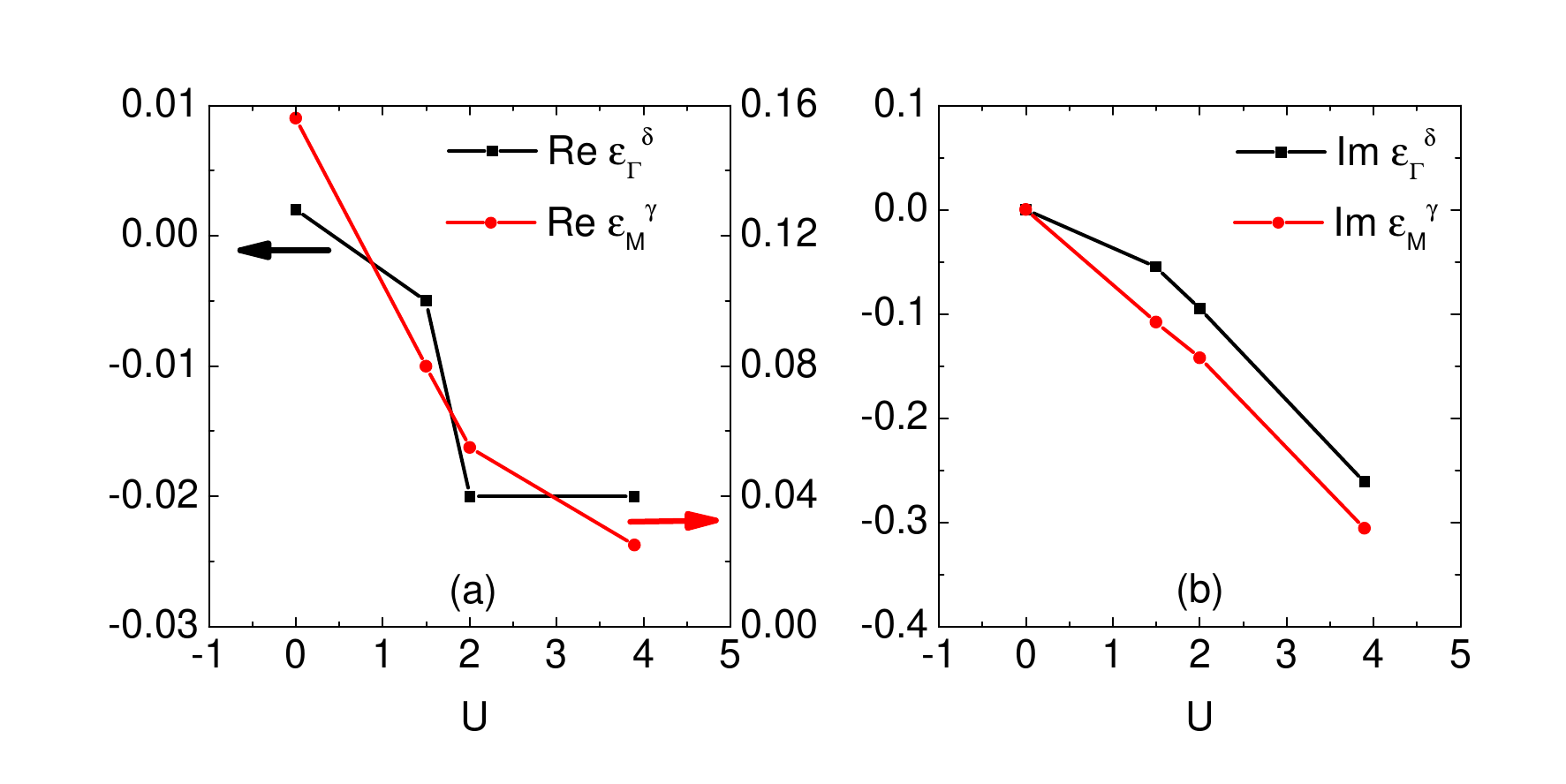}
 \caption{\label{self_energy} (a) The real part of the $\delta$ band energy at $\Gamma$ (black) and the $\gamma$ band energy at $M$ (red). (b) The imaginary part of the $\delta$ band energy at $\Gamma$ (black) and the $\gamma$ band energy at $M$ (red).}
\end{figure}

\begin{figure}
\includegraphics[width=1\linewidth]{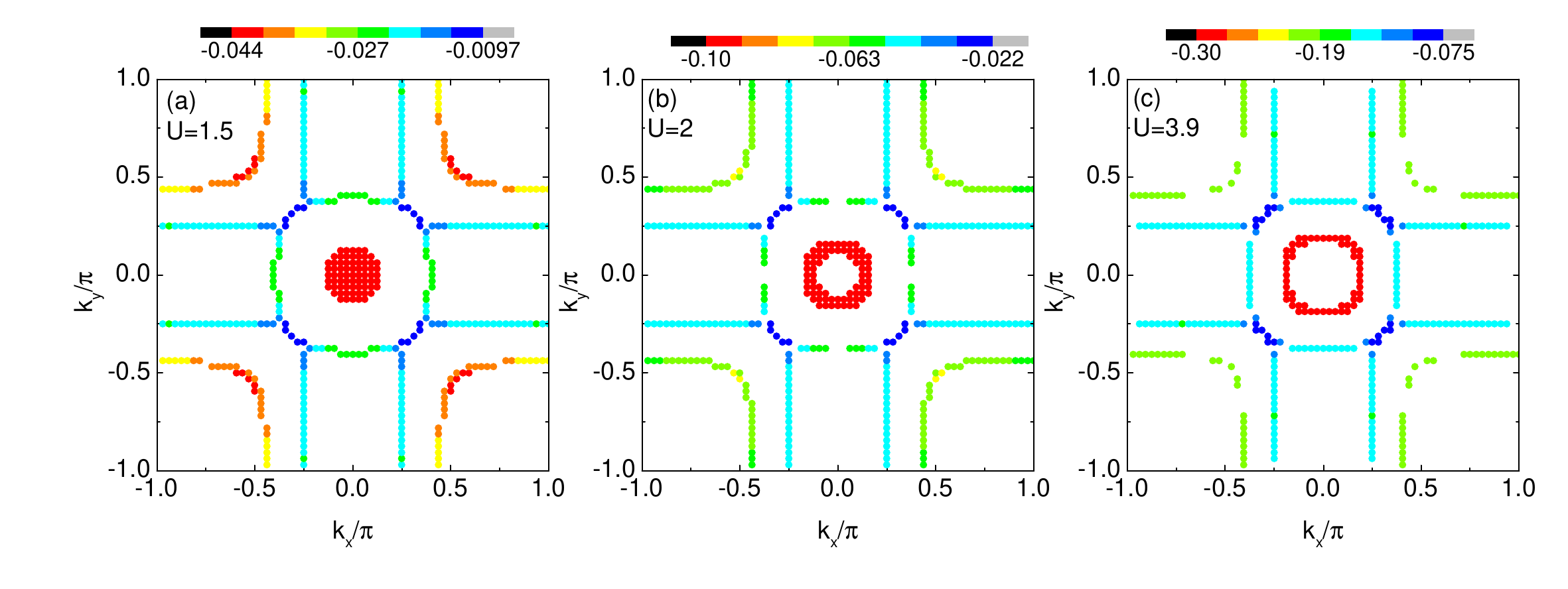}
 \caption{\label{fermi surface} The renormalized Fermi surface at $U=1.5$ (a), $U=2$ (b) and $U=3.9$ (c). The color scale represents the value of $\Im \varepsilon_{\mathbf{k}}^{s}(0)$.}
\end{figure}

\begin{figure}
\includegraphics[width=1\linewidth]{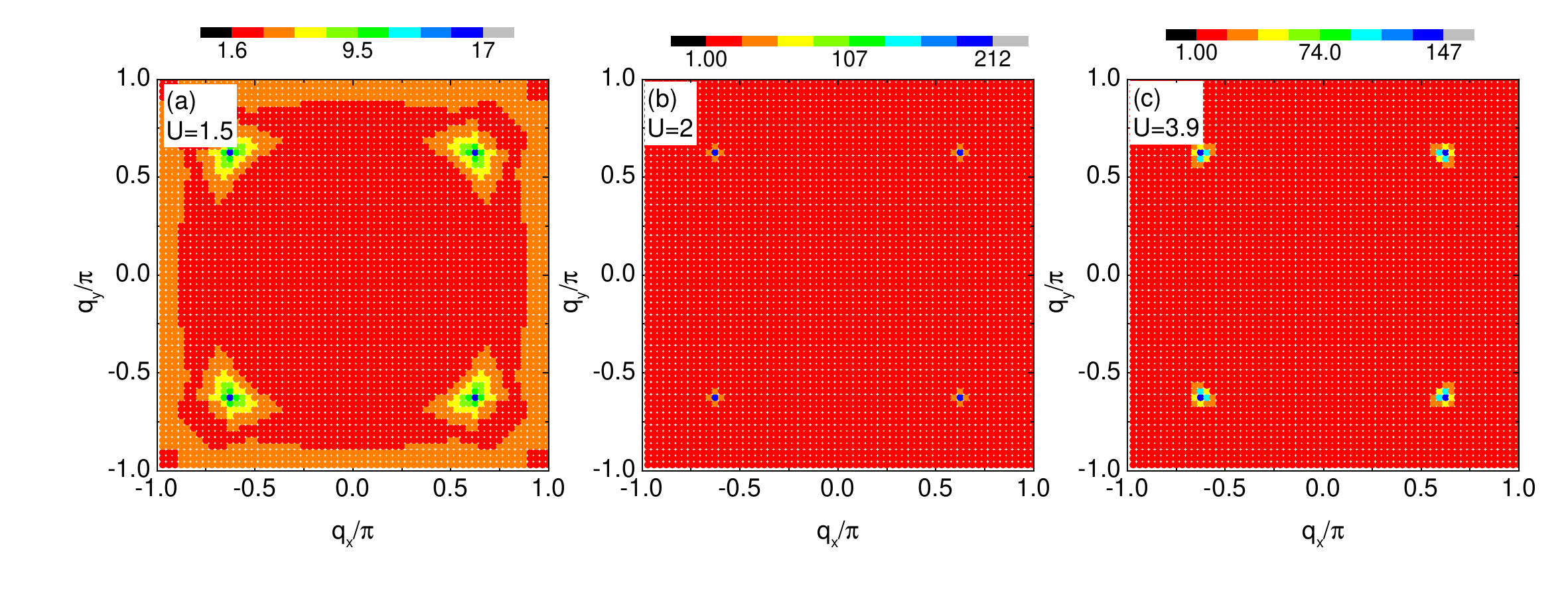}
 \caption{\label{spin_susceptibility} The largest eigenvalue of $\chi_s(\mathbf{q},i\omega_n=0)$ at $U=1.5$ (a), $U=2$ (b) and $U=3.9$ (c).}
\end{figure}

\begin{figure}
\includegraphics[width=1\linewidth]{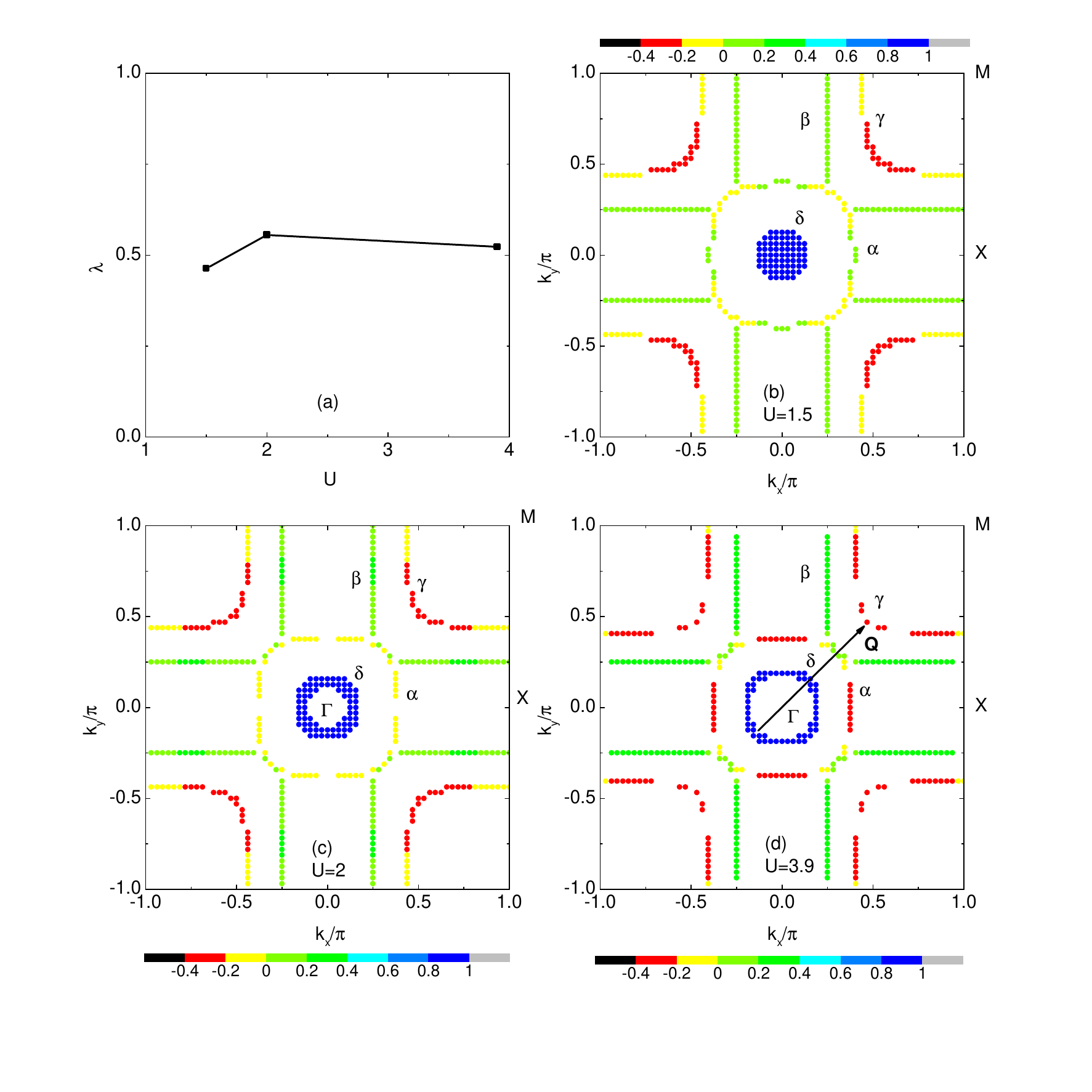}
 \caption{\label{lambda_and_pairing} (a) The largest positive eigenvalue $\lambda$ of Eq. (\ref{Eliashberg}) as a function of $U$. (b), (c) and (d) are the pairing function projected onto the Fermi surface at $U=1.5$, $2$, and $3.9$, respectively.}
\end{figure}

The above band renormalization can be explained by the normal self energy matrix $\Sigma(\mathbf{k},\omega+i\eta)$. It is complex symmetric (not hermitian) and shares the same structure as $M_{\mathbf{k}}$ in Eq. (\ref{h0}). We define
\begin{eqnarray}
\label{retard_Green's_function}
B_{\mathbf{k}}(\omega)&=&M_{\mathbf{k}}+\Sigma(\mathbf{k},\omega+i\eta)-\mu I.
\end{eqnarray}
Its eigenvalues are denoted as $\varepsilon_{\mathbf{k}}^{s}(\omega)$, with $s=\alpha,\beta,\gamma,\delta$. They are complex and can be obtained in the same way as Eq. (\ref{eigenvalues}). The energy $\omega$ that satisfies the  condition $\omega=\Re \varepsilon_{\mathbf{k}}^{s}(\omega)$ is the quasiparticle energy in band $s$ and the corresponding $-\frac{1}{\Im \varepsilon_{\mathbf{k}}^{s}(\omega)}$ represents the quasiparticle lifetime. In Fig. \ref{self_energy}(a) we show the real part of the band energy $\Re\varepsilon_{\mathbf{k}}^{\delta}$ at $\Gamma$ and $\Re\varepsilon_{\mathbf{k}}^{\gamma}$ at $M$, as a function of the interaction strength $U$. The $\delta$ band at $\Gamma$ evolves from slightly above the Fermi level at $U=0$, to $0.02$ eV below the Fermi level at $U=3.9$. In contrast, the $\gamma$ band at $M$ remains above the Fermi level, although it gets closer and closer as $U$ increases. On the other hand, the imaginary part of both the $\delta$ band at $\Gamma$ and the $\gamma$ band at $M$ becomes stronger as $U$ increases [see Fig. \ref{self_energy}(b)], therefore these bands are less coherent and even cannot be resolved there. The renormalized Fermi surface can be determined by $0=\Re \varepsilon_{\mathbf{k}}^{s}(0)$ and is shown in Fig. \ref{fermi surface}. As $U$ increases, the $\alpha$ and $\beta$ Fermi surfaces vary little, while the $\gamma$ and $\delta$ ones are enlarged. However, the imaginary part of the $\gamma$ and $\delta$ Fermi surfaces is strong, thus they are less coherent in the spectral function $A(\mathbf{k},0)$. In contrast, close to the diagonal direction, the imaginary part of the $\alpha$ and $\beta$ Fermi surfaces is very small, so the Fermi surfaces are sharp there.

Next we investigate the spin susceptibility $\chi_s(q)$ in Eq. (\ref{RPA}). For a given $q$, $\chi_s$ is a $16\times16$ hermitian matrix and we show the largest eigenvalue of $\chi_s(\mathbf{q},i\omega_n=0)$ in Fig. \ref{spin_susceptibility}. For all the interaction strength we studied, the spin susceptibility is peaked at $\mathbf{Q}=(\pm0.625\pi,\pm0.625\pi)$. By inspecting the matrix elements of $\chi_s(\mathbf{Q},i\omega_n=0)$, we find that $\sum_{l_1,l_2}(-1)^{l_1}(-1)^{l_2}\chi^{l_1zl_1z,l_2zl_2z}_s(\mathbf{Q},i\omega_n=0)$ is the strongest, which corresponds to the odd spin susceptibility $\chi^{o}=2(\chi_{\parallel}-\chi_{\perp})$ in the $z$ orbital, as defined in Ref. \onlinecite{oddspinsusceptibility}. Therefore, the $\mathbf{Q}$ peak in $\chi_s$ stems from the electron scattering between a $z$-orbital symmetric band and a $z$-orbital asymmetric one \cite{oddspinsusceptibility}, indicating a pivotal role played by the $3d_{z^{2}}$ orbital.

Finally we come to the superconducting pairing. The largest positive eigenvalue $\lambda$ of Eq. (\ref{Eliashberg}) is shown in Fig. \ref{lambda_and_pairing}(a), suggesting that in the range of $1.5\leq U\leq 3.9$, $\lambda$ does not change much and $\lambda\approx0.5$. The corresponding eigenvector $\phi(\mathbf{k},ip_n)$ then determines the superconducting pairing function in the layer, orbital, momentum and Matsubara frequency space. At $p_n=\pi T$, the function $\phi(\mathbf{k},i\pi T)$ can approximate the pairing at the Fermi level. In order to project the pairing function onto the Fermi surface, we define, similar to Eq. (\ref{retard_Green's_function}),
\begin{eqnarray}
\label{Hermitian}
B_{\mathbf{k}}^{'}&=&M_{\mathbf{k}}+\Re \Sigma(\mathbf{k},0+i\eta)-\mu I.
\end{eqnarray}
It is hermitian and its eigenvalues $\varepsilon_{\mathbf{k}}^{'s}=0$ with $s=\alpha,\beta,\gamma,\delta$ can approximately determine the Fermi surface. It amounts to neglecting the quasiparticle damping on the Fermi surface. The pairing function projected onto the Fermi surface can then be calculated as
\begin{eqnarray}
\label{pairing_on_FS}
\Delta_{\mathbf{k}}&=&Q_{\mathbf{k}}^{\dagger}\phi(\mathbf{k},i\pi T)Q_{\mathbf{k}},
\end{eqnarray}
where $Q_{\mathbf{k}}$ is the unitary matrix that diagonalizes $B_{\mathbf{k}}^{'}$. Figures \ref{lambda_and_pairing}(b), \ref{lambda_and_pairing}(c) and \ref{lambda_and_pairing}(d) show $\Delta_{\mathbf{k}}$ on the Fermi surface at $U=1.5$, $2$ and $3.9$, respectively. The pairing symmetry is the same, i.e., $s$-wave. On the $\alpha$ ($\beta$) Fermi surface, $\Delta_{\mathbf{k}}$ evolves from nodal at $U=1.5$, to nodeless and $\Delta_{\mathbf{k}}<0$ ($>0$) at $U=3.9$. On the other hand, on the $\gamma$ ($\delta$) Fermi surface, $\Delta_{\mathbf{k}}$ is nodeless and $\Delta_{\mathbf{k}}<0$ ($>0$). The largest magnitude of $\Delta_{\mathbf{k}}$ is on the $\delta$ and $\gamma$ Fermi surfaces. The reason is, these two Fermi surfaces can be connected by $\mathbf{Q}$, as denoted by the black arrow in Fig. \ref{lambda_and_pairing}(d). The $\gamma$ band is a $z$-orbital symmetric band, while the $\delta$ band is a $z$-orbital asymmetric one. Therefore, these two bands can fully utilize the spin fluctuation structure in Fig. \ref{spin_susceptibility}. Although the $\gamma$ and $\delta$ Fermi surfaces are highly damped, as shown in Fig. \ref{fermi surface}, however, as has been derived explicitly in Ref. \onlinecite{gaoyiunpublished}, the superconducting pairing is not restricted on the Fermi surface and can has a large magnitude even when the band does not form any Fermi surface. Thus by forming an anti-phase pairing function on them with large magnitude, the value of $\lambda$ can be effectively increased. In addition, $\phi(\mathbf{k},i\pi T)$ can be expressed as
\begin{eqnarray}
\label{fit}
\phi(\mathbf{k},i\pi T)&=&\begin{pmatrix}
f^{x}_{\mathbf{k}}&f^{'x}_{\mathbf{k}}&f^{xz}_{\mathbf{k}}&f^{'xz}_{\mathbf{k}}\\f^{'x}_{\mathbf{k}}&f^{x}_{\mathbf{k}}&f^{'xz}_{\mathbf{k}}&f^{xz}_{\mathbf{k}}\\
f^{xz}_{\mathbf{k}}&f^{'xz}_{\mathbf{k}}&f^{z}_{\mathbf{k}}&f^{'z}_{\mathbf{k}}\\f^{'xz}_{\mathbf{k}}&f^{xz}_{\mathbf{k}}&f^{'z}_{\mathbf{k}}&f^{z}_{\mathbf{k}}
\end{pmatrix}.
\end{eqnarray}
Here, $f^{x}_{\mathbf{k}}/f^{z}_{\mathbf{k}}$ is the intra-orbital and intra-layer pairing function in the $x/z$ orbital, while $f^{'x}_{\mathbf{k}}/f^{'z}_{\mathbf{k}}$ is the intra-orbital but inter-layer one. Furthermore, $f^{xz}_{\mathbf{k}}$ is the inter-orbital and intra-layer pairing function and $f^{'xz}_{\mathbf{k}}$ is the inter-orbital and inter-layer one. The
nonzero matrix elements can be approximately fitted to
\begin{eqnarray}
\label{fitcase1}
f^{x}_{\mathbf{k}}&=&-0.34,\nonumber\\
f^{'x}_{\mathbf{k}}&=&-0.15,\nonumber\\
f^{xz}_{\mathbf{k}}&=&-0.25(\cos k_x-\cos k_y)-0.12(\cos 2k_x-\cos 2k_y),\nonumber\\
f^{z}_{\mathbf{k}}&=&0.61+0.68(\cos k_x+\cos k_y)\nonumber\\
&+&0.35(\cos 2k_x+\cos 2k_y),\nonumber\\
f^{'z}_{\mathbf{k}}&=&-1.
\end{eqnarray}
From above we can see, $f^{z}_{\mathbf{k}}$ and $f^{'z}_{\mathbf{k}}$ have the largest magnitude, indicating that, the $3d_{z^{2}}$ orbital contributes not only significantly to the spin susceptibility, but also to the superconducting pairing. Transforming to the symmetric-asymmetric representation defined in Eq. (\ref{symmetric_antisymmetric}), the pairing function on the $z$-orbital asymmetric band can be approximated as $\Delta_{\mathbf{k}}=f^{z}_{\mathbf{k}}-f^{'z}_{\mathbf{k}}$. Around $\Gamma$, we have $f^{z}_{\mathbf{k}}\approx2.67$, which is in the opposite sign with respect to $f^{'z}_{\mathbf{k}}$ and leads to $\Delta_{\mathbf{k}}\approx3.67$ there. Since the $\delta$ band is $z$-orbital asymmetric around $\Gamma$, therefore, on the $\delta$ band around $\Gamma$, the superconducting pairing is the largest. This explains why in Figs. \ref{lambda_and_pairing}(b), \ref{lambda_and_pairing}(c) and \ref{lambda_and_pairing}(d), the pairing function has the largest magnitude on the $\delta$ Fermi surface.

\section{summary}
In summary, we have studied the electronic correlation effects in La$_{2.85}$Pr$_{0.15}$Ni$_2$O$_7$ film grown on SrLaAlO$_4$. The $\gamma$ and $\delta$ bands both move down with interaction, while their Fermi surfaces become highly damped and unresovable in experiments. In contrast, the $\alpha$ and $\beta$ Fermi surfaces do not change much with interaction and they are sharp enough to be detected experimentally. Since the $\gamma$ and $\delta$ bands are mostly consisted of the $3d_{z^{2}}$ orbital while the $\alpha$ and $\beta$ ones are from the $3d_{x^{2}-y^{2}}$ orbital, therefore the $3d_{z^{2}}$ orbital is more affected by interaction. The spin susceptibility always peaks at a wave vector connecting a $z$-orbital symmetric band and a $z$-orbital asymmetric one, i.e., the $\gamma$ and $\delta$ bands. The superconducting pairing symmetry is $s$-wave and robust against interaction. The largest pairing magnitude is always on the $\delta$ band. The band renormalization, spin susceptibility and pairing function all indicate that the $3d_{z^{2}}$ orbital plays the most important role in this material. In comparison, the same model has been previously studied by cluster dynamical mean-field theory (CDMFT) in Ref. \onlinecite{bilayermodel}. It found an upward shift of the $\delta$ band with correlation, in contrast to ours. In addition, the spin susceptibility in Ref. \onlinecite{bilayermodel} peaks at a different wave vector compared to ours. Despite these differences, both Ref. \onlinecite{bilayermodel} and the present work found a highly damped $\delta$ band and $\gamma$ Fermi surface, while the $\alpha$ and $\beta$ Fermi surfaces are sharp enough, especially close to the diagonal direction. Furthermore, the $s$-wave pairing symmetry is similar in Ref. \onlinecite{bilayermodel} and the present one, both of which show strong pairing in the $3d_{z^{2}}$ orbital. Since in CDMFT, the $\mathbf{k}$ dependence is neglected in $\Sigma(\mathbf{k},ip_n)$, therefore our work provides a complementary study to it and further verifies the robustness of the $s$-wave pairing symmetry.


\begin{thebibliography}{99}

\bibitem{discovery} H. Sun, M. Huo, X. Hu, J. Li, Z. Liu, Y. Han, L. Tang, Z. Mao, P. Yang, B. Wang, J. Cheng, D.-X. Yao, G.-M. Zhang, and M. Wang, Signatures of superconductivity near 80K in a nickelate under high pressure, Nature \textbf{621}, 493-498 (2023).

\bibitem{thinfilm1} E. K. Ko, Y. Yu, Y. Liu, L. Bhatt, J. Li, V. Thampy, C.-T. Kuo, B. Y. Wang, Y. Lee, K. Lee, J.-S. Lee,
B. H. Goodge, D. A. Muller, and H. Y. Hwang, Signatures of ambient pressure superconductivity in thin film La$_3$Ni$_2$O$_7$, Nature \textbf{638}, 935–940 (2025).

\bibitem{thinfilm2} G. Zhou, W. Lv, H. Wang, Z. Nie, Y. Chen, Y. Li,
H. Huang, W.-Q. Chen, Y.-J. Sun, Q.-K. Xue, and Z. Chen, Ambient-pressure superconductivity onset above 40 K in (La,Pr)$_3$Ni$_2$O$_7$ films, Nature \textbf{640}, 641–646 (2025).

\bibitem{arpes_la2prni2o7} B. Y. Wang, Y. Zhong, S. Abadi, Y. Liu, Y. Yu, X. Zhang, Y.-M. Wu, R. Wang, J. Li, Y. Tarn, E. K. Ko, V. Thampy, M. Hashimoto, D. Lu, Y. S. Lee, T. P. Devereaux, C. Jia, H. Y. Hwang, and Z.-X. Shen, Electronic structure of compressively strained thin film La$_2$PrNi$_2$O$_7$, arXiv:2504.16372.

\bibitem{arpes_la2.85pr0.15ni2o7} P. Li, G. Zhou, W. Lv, Y. Li, C. Yue,
H. Huang, L. Xu, J. Shen, Y. Miao, W. Song,
Z. Nie, Y. Chen, H. Wang, W. Chen, Y. Huang,
Z.-H. Chen, T. Qian, J. Lin, J. He, Y.-J. Sun,
Z. Chen, Q.-K. Xue, Angle-resolved photoemission spectroscopy of superconducting (La,Pr)$_3$Ni$_2$O$_7$/SrLaAlO$_4$ heterostructures, National Science Review, nwaf205 (2025).

\bibitem{arpes_(laprsm)3ni2o7} J. Shen, G. Zhou, Y. Miao, P. Li, Z. Ou, Y. Chen, Z. Wang, R. Luan, H. Sun, Z. Feng, X. Yong, Y. Li, L. Xu, W. Lv, Z. Nie, H. Wang, H. Huang, Y.-J. Sun, Q.-K. Xue, J. He, Z. Chen, Nodeless superconducting gap and electron-boson coupling in (La,Pr,Sm)$_3$Ni$_2$O$_7$ films, arXiv:2502.17831.

\bibitem{arpes_la2.79sr0.21ni2o7} W. Sun, Z. Jiang, B. Hao, S. Yan, H. Zhang, M. Wang, Y. Yang, H. Sun, Z. Liu, D. Ji, Z. Gu, J. Zhou, D. Shen, D. Feng, Y. Nie, Observation of superconductivity-induced leading-edge gap in Sr-doped La$_3$Ni$_2$O$_7$ thin films, arXiv:2507.07409.

\bibitem{s1} Y.-B. Liu, J.-W. Mei, F. Ye, W.-Q. Chen, and F. Yang, s$^{\pm}$-Wave Pairing and the Destructive Role of Apical-Oxygen Deficiencies in La$_3$Ni$_2$O$_7$ under Pressure, Phys. Rev. Lett. \textbf{131}, 236002 (2023).

\bibitem{s7} Q.-G. Yang, D. Wang, and Q.-H. Wang, Possible $s_\pm$-wave superconductivity in La$_3$Ni$_2$O$_7$, Phys. Rev. B \textbf{108}, L140505 (2023).

\bibitem{s8} C. Lu, Z. Pan, F. Yang, and C. Wu, Interlayer-Coupling-Driven High-Temperature Superconductivity
in La$_3$Ni$_2$O$_7$ under Pressure, Phys. Rev. Lett. \textbf{132}, 146002 (2024).

\bibitem{s9} H. Sakakibara, N. Kitamine, M. Ochi, and K. Kuroki, Possible High $T_c$ Superconductivity in La$_3$Ni$_2$O$_7$ under High Pressure through Manifestation of a Nearly Half-Filled Bilayer Hubbard Model, Phys. Rev. Lett. \textbf{132}, 106002 (2024).

\bibitem{s11} Y. Zhang, L.-F. Lin, A. Moreo, T. A. Maier, and E. Dagotto, Structural phase transition, $s_\pm$-wave pairing, and magnetic stripe order in bilayered superconductor La$_3$Ni$_2$O$_7$ under pressure, Nat. Commu. \textbf{15}, 2470 (2024).

\bibitem{s12} S. Ryee, N. Witt, and T. O. Wehling, Quenched Pair Breaking by Interlayer Correlations as a Key to Superconductivity in La$_3$Ni$_2$O$_7$, Phys. Rev. Lett. \textbf{133}, 096002 (2024).

\bibitem{sd1} Z. Liao, L. Chen, G. Duan, Y. Wang, C. Liu, R. Yu, and Q. Si, Electron correlations and superconductivity in La$_3$Ni$_2$O$_7$ under pressure tuning, Phys. Rev. B \textbf{108}, 214522 (2023).

\bibitem{sd2} H. Oh and Y.-H. Zhang, Type-II $t-J$ model and shared superexchange coupling from Hund's rule
in superconducting La$_3$Ni$_2$O$_7$, Phys. Rev. B \textbf{108}, 174511 (2023).

\bibitem{sd3} Z. Fan, J.-F. Zhang, B. Zhan, D. Lv, X.-Y. Jiang, B. Normand, and T. Xiang, Superconductivity in nickelate and cuprate superconductors with strong bilayer coupling, Phys. Rev. B \textbf{110}, 024514 (2024).

\bibitem{sd6} M. Bejas, X. Wu, D. Chakraborty, A. P. Schnyder, and A. Greco, Out-of-plane bond order phase, superconductivity, and their competition in the $t-J_\|-J_\bot$ model for pressurized nickelates, arXiv:2411.00269v1.

\bibitem{d1} K. Jiang, Z. Wang, and F.-C. Zhang, High-Temperature Superconductivity in La$_3$Ni$_2$O$_7$, Chinese Phys. Lett. \textbf{41}, 017402 (2024).

\bibitem{d3} F. Lechermann, J. Gondolf, S. B$\ddot{o}$tzel, and I. M. Eremin, Electronic correlations and superconducting instability in La$_3$Ni$_2$O$_7$ under high pressure, Phys. Rev. B \textbf{108}, L201121 (2023).

\bibitem{sd4} G. Heier, K. Park, and S. Y. Savrasov, Competing $d_{xy}$ and $s_\pm$ pairing symmetries in superconducting La$_3$Ni$_2$O$_7$: LDA+ FLEX calculations, Phys. Rev. B \textbf{109}, 104508 (2024).

\bibitem{sd5} D. K. Singh, G. Goyal, Y. Bang, Possible pairing states in the superconducting bilayer nickelate, arXiv:2409.09321v2.

\bibitem{d2} C. Xia, H. Liu, S. Zhou, and H. Chen, Sensitive dependence of pairing symmetry on Ni-$e_g$ crystal field splitting in the nickelate superconductor La$_3$Ni$_2$O$_7$, Nat. Commun. \textbf{16}, 1054 (2025).

\bibitem{junhuazhang} J. Zhang, Theory of spin-fluctuation induced superconductivity in iron-based superconductors, Doctoral dissertation Iowa State University Ames.

\bibitem{bilayermodel} C. Yue, J.-J. Miao, H. Huang, Y. Hua, P. Li, Y. Li, G. Zhou, W. Lv, Q. Yang, H. Sun, Y.-J. Sun, J. Lin, Q.-K. Xue, Z. Chen, W.-Q. Chen, Correlated electronic structures and unconventional superconductivity in bilayer nickelate heterostructures, arXiv:2501.06875.

\bibitem{scalapino} S. Graser, T. A. Maier, P. J. Hirschfeld, and D. J. Scalapino, Near-degeneracy of several pairing channels in multiorbital models for the Fe pnictides, New J. Phys. \textbf{11}, 025016 (2009).

\bibitem{kubo} K. Kubo, Pairing symmetry in a two-orbital Hubbard model on a square lattice, Phys. Rev. B \textbf{75}, 224509 (2007).

\bibitem{mahan} See Eqs. (3.5), (3.239) and (3.240), G. D. Mahan, Many-Particle Physics (Third Edition).

\bibitem{eliashberg} T. Takimoto, T. Hotta, and K. Ueda, Strong-coupling theory of superconductivity in a degenerate Hubbard model, Phys. Rev. B \textbf{69}, 104504 (2004).

\bibitem{power method} S. Andrilli and D. Hecker, Elementary Linear Algebra (Fifth Edition) (2016).

\bibitem{oddfrequency}  J. Linder and A. V. Balatsky, Odd-frequency superconductivity, Rev. Mod. Phys. \textbf{91}, 045005 (2019).

\bibitem{pade} H. J. Vidberg and J. W. Serene, Solving the Eliashberg Equations by Means of $N$-Point Pad$\acute{e}$ Approximants, J. Low Temp. Phys. \textbf{29}, 179-192 (1977).

\bibitem{oddspinsusceptibility} S. B$\ddot{o}$zel, F. Lechermann, J. Gondolf, and I. M. Eremin, Theory of magnetic excitations in the multilayer nickelate superconductor La$_3$Ni$_2$O$_7$, Phys. Rev. B \textbf{109}, L180502 (2024).
    
\bibitem{gaoyiunpublished} Y. Gao, Robust $s_\pm$-wave pairing in a bilayer two-orbital model of pressurized La$_3$Ni$_2$O$_7$ without the $\gamma$ Fermi surface, arXiv:2502.19840.

\end{thebibliography}
\end{document}